\documentclass[aps,prd,twocolumn,floatfix]{revtex4-2}

\usepackage{graphicx}
\newcommand{\be}{\begin{equation}}
\newcommand{\ee}{\end{equation}}
\newcommand{\bea}{\begin{eqnarray}}
\newcommand{\beas}{\begin{eqnarray*}}
\newcommand{\eea}{\end{eqnarray}}
\newcommand{\eeas}{\end{eqnarray*}}
\newcommand{\ba}{\begin{array}}
\newcommand{\ea}{\end{array}}
\begin{document}
%\markboth{Abdel P\'erez-Lorenzana}{Small $\theta_{13}$ and solar neutrino oscillation parameters from $\mu-\tau$ symmetry}

\title{Small $\theta_{13}$ and solar neutrino oscillation parameters from $\mu-\tau$ symmetry}

\author{Abdel P\'erez-Lorenzana}
\email{ aplorenz@fis.cinvestav.mx }

\affiliation{
%\address{
Departamento de F\'{\i}sica. Centro de Investigaci\'on y de Estudios Avanzados del I.P.N.,\\
Apdo. Post. 14-740, 07000, Mexico City, Mexico. }
%\email{aplorenz@fis.cinvestav.mx }

%\maketitle

\begin{abstract}
Exchange $\mu-\tau$ symmetry in the effective Majorana neutrino mass matrix does predict a maximal mixing for atmospheric neutrino oscillations asides to a null mixing  that cannot be straightforwardly identified with reactor neutrino oscillation mixing, $\theta_{13}$, unless a specific ordering is assumed for the mass eigenstates. Otherwise, a non zero value for $\theta_{13}$ is predicted already at the level of an exact  symmetry.
In this case, solar neutrino mixing and scale, as well as the correct atmospheric mixing,  arise from the breaking of the symmetry.
I present a mass matrix proposal for normal hierarchy that realizes this scenario, where the smallness of $\tan\theta_{13}$  is naturally given by the parameter  $\epsilon\sim\sqrt{\Delta m^2_{sol}/\Delta m^2_{ATM}}$ and the solar mixing is linked to the smallness of $\Delta m^2_{sol}$.  The proposed matrix remains stable under renormalization effects and it also allows to account for CP violation within the expected region without further constrains.

%\keywords{Neutrino masses and mixings, Flavor symmetries}
\end{abstract}
%\date{\today}
\maketitle

%\pacs{}

%%%%%%%%%%%
%%%%%%%%%%%%%%%
\section{Introduction}
%%%%%%%%%%%%%%%%
Although the Standard Model (SM) of particle physics was built on the basis of massless neutrinos, a plethora of data collected by neutrino oscillation experiments, using solar, atmospheric, reactor and accelerator neutrino sources, had provided uncontroversial evidence that neutrinos are actually massive and light~\cite{nuoscPDG}. Most observed oscillation phenomena is well explained if standard  weak neutrino flavors are not the actual mass egienstates, but rather a combination of the last, such that $\nu_{\ell L}= U_{\ell i}~\nu_{iL}$, for $\ell = e,\mu,\tau$ and $i=1,2, 3$. 

The  mechanism that provides neutrino masses is still unknown, yet, Majorana neutrinos seem the most natural choice, since the simple addition of  missing right handed neutrinos to the SM would give an explanation to the lightness of the active neutrinos, through the seesaw mechanism~\cite{seesaw}. In any case, effective Majorana mass terms written in the weak basis,  
$(M_\nu)_{\ell\ell^\prime} \bar{\nu}_{\ell L}^c \nu_{\ell^\prime L}$, 
should be non diagonal. Diagonalization of these would proceed through an orthogonal transformation with the unitary mixing matrix $U$, so that 
$M_{diag} =  U^T M_\nu U$, where 
$M_{diag} =  \text{Diag}\{ m_1 ,m_2 , m_3 \}$, with $m_i$ the real masses.

The $U$ matrix can be written in terms of three complex rotations by using Pontecorvo-Maki-Nakagawa-Sakata (PMNS)~\cite{pontecorvo,MNS} parameterization, where  
$U = V\cdot K$, and 
%\begin{widetext}
%\begin{equation}
%\label{PMNS}
$$
V=  \left( \begin{array}{ccc} 
c_{12} c_{13} & s_{12}c_{13} & z \\
- s_{12} c_{23} - c_{12}s_{23}\bar{z}  & 
c_{12}c_{23} - s_{12}s_{23}\bar{z} & s_{23}c_{13} \\
s_{12}s_{23} - c_{12}c_{23}\bar{z} & 
- c_{12}s_{23} -c_{23}s_{12}\bar{z} & c_{23}c_{13} 
\end{array} \right)~, 
$$
%\end{equation*} 
%\end{widetext}
with $c_{ij}$ and $s_{ij}$ standing for $\cos \theta_{ij}$ and  
$\sin \theta_{ij}$, respectively,  of  the mixing angles given as 
$\theta_{12}$, $\theta_{13}$, and $\theta_{23}$. Here, $z= s_{13}e^{-i\delta_{CP}}$, where $\delta_{CP}$ is the
Dirac $CP$ phase, whereas $K$ is a diagonal matrix 
containing two Majorana phases which do not contribute to neutrino 
oscillations. In the basis where charge lepton masses are diagonal, same that we assume hereafter, $U$ also  describes the mixings involved in charged weak currents, $W_\mu \bar \ell_L \gamma^\mu U_{\ell i}\nu_{iL}$.

Neutrino oscillation experiments are actually sensible to above mixings and to mass squared differences, $\Delta m^2_{ij} = m_i^2 - m_j^2$, identified as solar $\Delta m^2_{sol} = \Delta m^2_{21}$ and atmospheric
$\Delta m^2_{ATM}=|\Delta m^2_{31}|$ scales. Due to matter effects within the sun, it is know that $\Delta m^2_{21}>0$, whereas the sign of $\Delta m^2_{31}$ and therefore the hierarchy of masses is still unknown. If $m^2_3>m^2_1$ we would have a normal hierarchy (NH), otherwise it is called inverted.

Global fits of oscillation data with all three neutrinos~\cite{nuglobal} indicate that in NH, for instance, at one sigma level,
$\Delta m_{sol}^2 = 7.42^{+0.21}_{-0.20}\times 10^{-5}~eV^2$ and 
$\Delta m_{ATM}^2 = 2.517^{+0.026}_{-0.028}\times 10^{-3}~eV^2$, 
$\sin^2\theta_{12} = \sin^2\theta_{sol}=0.304^{+0.012}_{-0.012}$, 
$\sin^2\theta_{23}= \sin^2\theta_{ATM}=0.573^{+0.016}_{-0.020}$, and 
$\sin^2\theta_{13}= 0.02219^{+0.00062}_{-0.00063}$, corresponding to reactor oscillation mixings. On the other hand, $\delta_{CP}$ should be within the interval $[120^o,369^o]$ at three sigma level, which still allows for non CP violation ($\delta_{CP}=180^o$ or $360^o$).

In the attempt to understand such parameter pattern, flavor symmetries had long been advocated as a possible explanation of the observed mass hierarchies and mixings among fundamental fermions. Amid the possibilities, $\mu-\tau$ exchange symmetry stands out since it naturally appears in $M_\nu$ when maximal $\theta_{23}$ and zero $\theta_{13}$ are taken for $U$, and a large number of studies had been dedicated to explore this symmetry and its possible realization in extended fermion mass models~\cite{mutau,others,mtreviews}. However, all such studies always assume that a null $\theta_{13}$ is a necessary outcome of $\mu-\tau$ symmetry and its non zero observed value as the signature of the breaking of the symmetry. Also, this approach leaves out solar mixing, offering no understanding for its observed value whatsoever.

It turns out, however, that above claims are not compulsory. 
As I will discuss along this paper, the actual signature of exact $\mu-\tau$ symmetry is solely: 

(i) the maximality of $\theta_{23}$, 

(ii) a zero Dirac CP phase, and 

(iii) the existence of an null mixing.\\
Identifying  the last with $\theta_{13}$ is just a particular case subjected to a specific ordering of mass eigenstates. As a matter of fact, there is a possible ordering where $\theta_{13}$ arises with a non zero value in the exact symmetry limit. In such a scenario it is rather $\theta_{12}$ which results null. Although such an outcome seems unwanted, given the observed large value of solar mixing, the right value can be obtained from the breaking of $\mu-\tau$ symmetry, to the cost of starting with a degenerate spectrum in the one-two sector before considering symmetry  breaking  corrections. With this mechanism, the emergence of the solar mixing gets attached to the origin of $\Delta m^2_{sol}$.

To support these claims, I shall present a mass matrix structure for normal hierarchy  where both $\tan\theta_{13}$ and the breaking of $\mu-\tau$ are governed by a single parameter, the ratio $\sqrt{\Delta m^2_{sol}/\Delta m^2_{ATM}}$. As I will show, the proposed $M_\nu$ successfully predicts the oscillation mass scales, a small reactor mixing, an atmospheric mixing that is larger than $\pi/4$, and a large solar mixing,  all consistent with observed values. Besides, it also allows to incorporate CP violation without losing above features.

The discussion is organized by starting with a brief review of the general $\mu-\tau$ symmetric predictions to state the basis of the case that will be considered along the paper. Then, the texture under study and their predictions shall be introduce in section III. Stability under renormalization effects and the exploration of CP violation are addressed in sections IV and V, respectively, and the paper is closed with some remarks and conclusions.

%%%%%%%%%%%%%%%
\section{masses and mixings from $\mu-\tau$ symmetry}
%%%%%%%%%%%%%%%%

Assuming $\mu-\tau$ symmetry in  the  Majorana neutrino mass matrix means that the matrix elements, $m_{\ell,\ell^\prime}$ should comply with the identities $m_{e\mu} = m_{e\tau}$ and  $m_{\mu\mu} = m_{\tau\tau}$. 
On the other hand, in order to determine the mixing matrix we should consider the hermitian squared matrix $H = MM^\dagger$, which would explicitly exhibit $\mu-\tau$ symmetry too. $H$ can in general be written as
\be 
H= \left(
\ba{ccc}
\alpha     & \omega & \omega \\
\bar\omega & \beta  & \rho \\
\bar\omega & \rho   & \beta  
\ea
\right)
\label{H}
\ee
where $\alpha = |m_{ee}|^2 + 2 |m_{e\mu}|^2$, $\beta = |m_{\mu\mu}|^2 + |m_{e\mu}|^2 + |m_{\mu\tau}|^2$, and $\rho= 2Re(m_{\mu\mu}\bar m_{\mu\tau}) + |m_{e\mu}|^2$, with  the only complex component being  $\omega=m_{ee}\bar m_{e\mu} + m_{e\mu}(\bar m_{\mu\mu} + \bar m_{\mu\tau})$. The mixing matrix $U$ should then diagonalize $H$ through the unitary transformation $U^T HU^* = M_{diag}^2$. Thus,  the complex conjugated orthonormal eigenvectors of $H$ do correspond to the columns of the mixing matrix. Furthermore, given that the characteristic polynomial can be factored as
$p_H(\lambda) = [(\beta -\rho) - \lambda][(\alpha - \lambda)(\beta +\rho - \lambda) - 2|\omega|^2]$,
the squared eigenmasses are $\lambda_0 = \beta-\rho$
and 
\be
\lambda_{\pm} = \frac{1}{2}\left[\alpha + \beta + 
\rho \pm \sqrt{\left(\alpha - (\beta+\rho)\right)^2 + 8|\omega|^2}\right]~.
\ee
The corresponding eigenvector for $\lambda_0$ is 
\be 
\nu_0 = \frac{1}{\sqrt{2}}\left(\ba{r} 0\\ -1\\ 1 \ea \right)~. 
\ee
Clearly, this implies $\mu$ and $\tau$ maximal mixing as the content of such eigenstate, which means a maximal $\theta_{23}$ on PMNS matrix, regardless of mass ordering. Moreover, and yet more interestingly, other mixings can be pined down right out of this unique eigenstate, as I discus next.

By construction, $\lambda_+>\lambda_-$, thus, mass ordering is defined by the sole location of $\lambda_0$ within the spectrum. Either, (a) $\lambda_0>\lambda_+$, (b) $\lambda_+> \lambda_0>\lambda_-$, or (c) $\lambda_->\lambda_0$. Nonetheless, neutrino hierarchy would also be associated to the value of $\Delta\lambda = \lambda_+-\lambda_-$, which could either be about solar or atmospheric scale. Because of this, all  above possible orderings may result phenomenologically viable, and so, $\lambda_0$ could play the role of any $m_i^2$ in the spectrum, with $\nu_0$  realizing any of the columns on PMNS mixing matrix. Hence,  $\mu-\tau$ symmetry would have, besides a maximal $\theta_{23}$, any of the following outcomes. 

(i) For $\lambda_0= m_3^2$ one gets $z=0$, and thus a null $\theta_{13}$ and non Dirac CP phase.  This is the case implicitly assumed in most $\mu-\tau$ models studied so far in the literature.

(ii) For $\lambda_0= m_2^2$ one rather gets $\sin\theta_{12}= 0$, whereas,

(iii) $\lambda_0= m_1^2$, implies $\cos\theta_{12} = 0$. 

On last two cases it is obvious that orthogonality of the eigenstates means that $|z|\neq 0$, and hence a non zero $\theta_{13}$. A calculation  gives the general formula 
\be
\tan\theta_{13}=\frac{|\rho+\beta-m_3^2|}{\sqrt{2}\, |\omega|}~.
\label{symmtheta13}
\ee
Also, it is not difficult to check that in latter cases the Dirac CP phase becomes $\delta_{CP} = \arg(\omega)$, but it can always be factored out from the mixing matrix and absorbed within the Majorana phases. Thus, it can consistently be taken as zero at this level. Note that if $\lambda_0$ is degenerated with any other eigenvalue, scenarios (ii) and (iii) reduce to a single one where $\theta_{12}=0$. 

None of the above scenarios is consistent with data, yet closeness to maximal mixing in atmospheric oscillations suggest that $\mu-\tau$ could be treated as a slightly broken symmetry. Indeed, as the many previous studies show, symmetry breaking  can eventually explain the non zero value of $\theta_{13}$ in the first scenario. I will not elaborate on such a case any further in here. Rather, I shall address the more interesting question raised by the other possible scenarios. In both of them, a small reactor mixing comes at the symmetric limit  as a consequence of the interplay of matrix elements. Therefore, one could naturally expect that a small breaking of $\mu-\tau$ would provide small corrections to $\theta_{23}$ and $\theta_{13}$, so to explain the experimental values.
Nevertheless, given that the  input value from the symmetric limit for $\theta_{12}$  is either $0$ or $\pi/2$, it might appear challenging to understand from the same perspective the rather large measured value of solar neutrino mixing. It turns out that such is not the case. As a matter of fact, a large $\theta_{12}$ mixing can be obtained from zero mixing through perturbations, provided the initial spectrum is degenerate ($\Delta m^2_{12} = 0$). 
To illustrate this point, consider the following toy matrix,
\be
A= \left (\ba{cc} m &\varepsilon\\ \varepsilon & m \ea \right)~.
\ee
Clearly, at the limit when $\varepsilon=0$, there is no mixing at all, and states are degenerated. However, regardless of its value, as soon as we take $\varepsilon\neq 0$, the mixing becomes maximal, meaning $\tan\theta = 1$, which happens also regardless of the value of $m$.  Accommodating a value in between for $\theta$ becomes a matter of  fixing  $A$. Besides, the same perturbation also breaks the degeneracy,  introducing a mass gap $\Delta m^2 = 4m\varepsilon$. In this line of thought, 
given that $\Delta m_{sol}^2$ is the smallest of the scales, it comes natural to think of it as indicating an initial degeneracy on the $m_1$-$m_2$ sector that sources the large solar mixing when lifted. That is indeed possible, and as I discuss next, one can use a realization of this mechanism in the context of three neutrinos to get a working structure for $M_\nu$ .

%%%%%%%%%%%%%%%
\section{Generating solar oscillation parameters}
%%%%%%%%%%%%%%%%

For the rest of our discussion let us consider the following mass matrix structure for normal neutrino hierarchy,
\be \label{texture}
M_\nu= \left (\ba{ccc} 
d\epsilon^2 &\epsilon &a\epsilon\\ 
\epsilon & b & c \\
a\epsilon & c & 1
\ea \right)m_0~.
\ee
that explicitly realizes $\mu-\tau$ symmetry for $a=b=1$.
For the overall neutrino scale one naturally expects $m_0\sim\sqrt{\Delta m^2_{ATM}}$, and  I also assume 
\be\label{epsilon}
\epsilon\sim\sqrt{{\Delta m^2_{sol}}/{\Delta m^2_{ATM}}}= 0.172~.
\ee
The hierarchies within $M_\nu$ could be realized in flavor models with an approximated $U(1)_{L_e}$ symmetry (additional to $\mu-\tau$), where $\epsilon$ encodes the amount of breaking of $L_e$, but I will not pursue such models here.
Upon weak neutrino phase redefinition, one can take $b$ and $\epsilon$ as real numbers, while $a$, $c$ and $d$ remain complex. Their phases would  non trivially combine to generate observable phases. However, in order to explain the way neutrino oscillation parameters arise from above ansatz, while keeping the discussion simple,  I will assume for the moment non CP violation, and take all matrix elements as real. I shall come back to the complex case latter on. 

Note that for $a=b=c=d=1$, $M_\nu$ reduces to 
\be\label{M0}
 \left (\ba{ccc} 
\epsilon^2 &\epsilon &\epsilon\\ 
\epsilon & 1 & 1 \\
\epsilon & 1 & 1
\ea \right)m_0~,
\ee
which exhibits $\mu-\tau$ symmetry. Furthermore, its eigenvalues become
 $m_1 = m_2 = 0$ and $m_3 = (2+\epsilon^2)m_0$, which suggests that $m_0^2\sim \Delta m^2_{ATM}/4$, a value that is consistent with NH, but for the fact that non solar neutrino oscillations scale exist at this level. It is easy to see that this does correspond to the scenario at hand, since at this level $\sin\theta_{12}=0$,  whereas for the reactor mixing one gets \be 
 \tan\theta_{13} = \frac{\epsilon}{\sqrt{2}}~.
 \ee
 Notice that if $\epsilon$ gets the exact value in RHS of Eq.~(\ref{epsilon}), one gets  the $\mu-\tau$ symmetric prediction 
 $\tan\theta_{13}^{\tiny{(s)}} = 0.1215$, that results a bit smaller than the central measured value, $\tan\theta_{13}= 0.1506$.  Of course, this can be solved with  a bit larger value for $\epsilon$. Its actual  value, though, would emerge once $\mu-\tau$ gets broken.

As already stated above, generating the missing solar neutrino parameters requires to work on the broken symmetry case, although, to avoid large corrections to $\theta_{ATM}$ and $\theta_{13}$ one has to do it perturbatively, in terms of $\epsilon$.
To such a purpose, let us come back to $M_\nu$ given in Eq.~(\ref{texture}),  and subject its parameters to the convenient condition,
\be
b= 1+\delta_b \epsilon~,\quad\text{and}\quad c= 1- \delta_c \epsilon^2~,
\ee
with $d$ and $\delta_{b,c}$ order one parameters. 
On the other hand, $a$ should range within order one to $10^{-1}$. This parameter should be set later on to a value that properly fixes the  neutrino observable outcomes. After some algebra, a perturbative diagonalization of the so given $M_\nu$ shows that the predicted mixings, at leading order in $\epsilon$, are
\be
%\nonumber
\tan\theta_{ATM}\approx 1 + \frac{\delta_b}{2}\epsilon~,\quad
\tan\theta_{13}\approx \frac{\epsilon}{2\sqrt{2}}(1+a)~,
\label{thetaatm}
\ee
and
\be
\tan\theta_{sol}\approx \frac{2\sqrt{2}(1-a)}{\delta_b + \sqrt{\delta_b^2 + 8(1-a)^2}}
\label{tansol}
\ee
Above expressions show that, indeed, the proposed mass matrix provides a correction on the atmospheric mixing in the right direction, tending to increase the angle above $\pi/4$, whereas adding a small factor correction to reactor mixing, as given in the $\mu-\tau$ limit, provided $\delta_b>0$ and $a$  is a small number.
More importantly, the given $M_\nu$ predicts a  solar mixing that does not depend on the expansion parameter, $\epsilon$, as expected, but links its value with the atmospheric mixing through $\delta_b$ and to reactor mixing through $a$, the actual $\mu-\tau$ breaking parameters.

Additionally, the perturbation does lift the degeneracy on the lightest sector, and one gets 
\be\label{deltasol}
\Delta m^2_{sol}\approx \epsilon^2 m_0^2\left(\frac{\delta_b}{4}\right) \sqrt{\delta_b^2 + 8(1-a)^2} ~,
\ee
which is given in terms of the very same parameters that provide solar mixing.
On the other hand, for the atmospheric scale one gets, 
\be\label{deltaatm}
\Delta m^2_{ATM}\approx (4 + 2\delta_b\epsilon+\frac{1}{8}C\epsilon^2)m_0^2
\ee
with 
$C= 8(1+a)^2 - 4(1-a)^2 +5\delta_b^2 - 32\delta_c - \delta_b\sqrt{\delta_b^2+8(1-a)^2}$. 

The set of equations (\ref{thetaatm}) - (\ref{deltaatm})  provides a unique solution for the involved mass matrix parameters. Observed mixings set the numerical values for $\delta_b$, $a$ and $\epsilon$, whereas oscillation scales fix the values for $\delta_c$ and $m_0$. A numerical solution, with central values of the oscillation neutrino parameters, gives
$a=0.237 $, $\delta_b = 0.92$,  $\epsilon =0.34$,
$\delta_c = 5.49$, and  $m_0 =0.034~eV$~. 
This outcomes validate our approximation. Notice that $d$ has no relevance in this calculation. It only enters at the $\epsilon^3$ order.
Of course, small deviations on above  values would still provide neutrino oscillation parameters  well within accepted experimental ranges. 

The small value of $a$ signals that the breaking of $\mu-\tau$ in $m_{e\tau}$ is the strongest one. The correction should amount to reduce an entrance of order $\epsilon$, as in Eq.~(\ref{M0}), down to a value about $\epsilon^2$. Yet, this is still a small correction with respect to the overall $m_0$ scale. One has to keep in mind that this conclusion holds for the non CP violationg case. As it is discussed below, the addition of CP phases would also provide a way to overcome the condition of an small $a$, providing the possibility to account for the neutrino observables even with an small amount of breaking  for $\mu-\tau$ symmetry.

%%%%%%%%%%%%%%%
\section{Stability under RGE evolution}
%%%%%%%%%%%%%%%%
As it s well known~\cite{RGE}, renormalization of neutrino mass may substantially alter mass textures and their predictions when these last heavily relay on the degeneracy of the eigenvalues. 
(For a recent discussion about the integral solutions to RGEs for neutrino masses and flavor mixing parameters  in both the Majorana and Dirac cases see Ref.~\cite{zhang}.)
At first glance this could be considered a matter of concern in the present case, where the mass matrix in Eq.~(\ref{texture}) is assumed to be realized in a flavor model that includes $\mu-\tau$ symmetry at some higher energy $\Lambda$,
and it has to be addressed by explicitly including the effect or renormalization  to run the mass matrix elements down to a lower scale, $\Lambda_0$, usually taken as the electroweak scale. 
Nonetheless, it should be stressed that even though we started the discussion by requiring degeneracy on two neutrino states, the proposal to understand the solar mixing does also require to break such degeneracy down by opening the mass gap associated to solar neutrino scale, $\Delta m^2_{sol}$. Thus, the texture (\ref{texture}), expected to emerge from the  flavor model at the scale $\Lambda$, does not produce degenerated neutrinos, but rather a hierarchical spectrum. Henceforth, one should not expect any mayor distortion to the proposed solution to neutrino masses and mixings from renormalization group equations (RGE). This is explicitly shown next. 

At one-loop order, the low energy effective neutrino  mass is defined from the high energy one by the scaling induced by the RGE, as follows~\cite{RGE},
\be
M_\nu(\Lambda_0) = I_\gamma~ {\cal I}M_\nu(\Lambda) {\cal I}~,
\ee
where the diagonal matrix ${\cal I} = Diag\{I_e,I_\mu,I_\tau \}$, whose entries are given by the integral expression
\be
I_\ell = exp\left[-\frac{f}{16\pi^2}\int_{\ln\Lambda_0}^{\ln\Lambda}y_\ell^2~dt\right]~,
\ee
whereas the overall scaling factor is written as
\be
I_\gamma = exp\left[-\frac{1}{16\pi^2}\int_{\ln\Lambda_0}^{\ln\Lambda}\gamma(t)~dt\right]~.
\ee
In above, considering the Standard Model, the scale function
$\gamma =-3g_2^2 + \lambda + 6 y_t^2 $, where  $g_2$ stands for the $SU(2)_L$ guage coupling, $\lambda$ for the Higgs self coupling and  $y_{t,\ell}$  for the top and lepton Yukawa couplings, whereas $f = -3/2$.

The overall factor $I_\gamma$ contributes by only  redefining the overall neutrino scale,  and thus, it has to be considered to fix the correct value that should provide the right neutrino oscillation scales at low energy. It has no relevance for renormalization of the mixing angles, though.  

Flavor dependent corrections, $I_\ell$, on the other hand, break by themselves $\mu-\tau$ symmetry and constitute an additional source that alters the mixings defined at the symmetric limit. However, since lepton Yukawas are rather small and hierarchical, with $y_e<<y_\mu<<y_\tau$, it is expected that $I_\ell$ would be close to unity, with the main contributions coming from tau lepton. In general, one can write $I_\ell=e^{r_\ell}\approx 1 + r_\ell$, where
\be
r_\ell \approx \frac{3}{32\pi^2}y^2_\ell(\Lambda_0)\ln\left(\frac{\Lambda_0}{\Lambda}\right)~.
\ee
By assuming  $\Lambda = \Lambda_{GUT}\sim 10^{16}~GeV$ and $\Lambda_0$ the electroweak scale, one gets that $r_\tau\sim 10^{-5}$, whereas $r_{\mu}\sim 10^{-7}$ and $r_e\sim 10^{-8}$.

Under this prescription, taking $M_\nu(\Lambda)$ as the mass matrix given in Eq.~(\ref{texture}), the effective mass matrix at low energy becomes
\be
M_{\nu}(\Lambda_0) = 
\left (\ba{ccc} 
d_R\epsilon_R^2 &\epsilon_R &a_R\epsilon_R\\ 
\epsilon_R & b_R & c_R \\
a_R\epsilon_R & c_R & 1
\ea \right)m_{0,R}~,
\ee
where the index $R$ refers to the renormalized quantities, given as
\bea
 a_R = \left(\frac{I_\tau}{I_\mu}\right)a~, &\qquad &
b_R = \left(\frac{I_\mu}{I_\tau}\right)^2 b~, \\
c_R = \left(\frac{I_\mu}{I_\tau}\right) c~, &&
d_R = \left(\frac{I_\tau}{I_\mu}\right)^2 d~, 
\eea
\be
\epsilon_R = \left(\frac{I_eI_\mu}{I_\tau^2}\right)\epsilon~,
\ee
and where the overall neutrino scale $m_{0R} = (I_\gamma I_\tau^2)\,m_0$. 

Due to the smallness of $r_{e,\mu}$ with respect to $r_\tau$, it is enough to consider only the effect of the tau lepton corrections in $M_\nu(\Lambda_0)$.  In this approach, it is straightforward to see that the renormalized parameters involved in the mixings just get an small factor correction, of order ${\cal O} (10^{-5})$, with respect to those defined at high energy. Explicitly, one gets 
$a_R \approx (1+r_\tau)a$, 
$b_R \approx (1-2r_\tau) b$, 
$c_R \approx (1-r_\tau) c$,  
$d_R \approx (1+2r_\tau) d$, and $\epsilon_R \approx (1-2r_\tau)\epsilon$. 
As expected, these effect should not alter the texture itself, nor the predictions of the theory, in any sensible way. As a matter of fact, the  formulae (\ref{thetaatm}) and (\ref{tansol}) for the mixing angles would still be valid for the renormalized parameters. This allow to express, to the lowest order, the renormalized mixings  as
\bea
\tan\theta_{ATM}^R &\approx& \tan\theta_{ATM} - r_\tau~,\\
\tan\theta_{13}^R&\approx&\tan\theta_{13}\left(1-\frac{2+a}{1+a}r_\tau\right) ~,
\eea
and
\be
\tan\theta_{sol}^R = \tan\theta_{sol}\left(1+\Delta R\right)~,
\ee
where, to smplify matters, I have used a short hand notation for 
$\Delta R = \tan\theta_{sol}\left(1+(1+x^2)^{-\frac{1}{2}}\right)\Delta s$, with  $x=\delta b/\sqrt{8}(1+a)$ and 
\be
\Delta s = \frac{(2-a)\delta_b-2b(1-a)}{(1-a)^2}\frac{r_\tau}{\epsilon}~.
\ee
This correction term remains small,  about $\Delta R\sim 10^{-5}$.
Hence, all corrections induced on the mixing angles by running under RGE do came out to be very small.

%%%%%%%%%%%%%%%
\section{CP violation}
%%%%%%%%%%%%%%%%

Setting back CP phases in $M_\nu$ would slightly modify above formulae for mass scales and mixings, and provide observable phases. As mentioned, relevant phases arise on $a$, $c$ and $d$ parameters, called $\phi_{a,c,d}$ respectively. 
In order to maintain the predictions given above one should now assume  $|b|= 1+\delta_b\epsilon$ and $|c|= 1- \delta_c\epsilon^2$,  to get 
$$
\tan\theta_{ATM}\approx 1+\frac{\delta_b\epsilon}{2|\cos\phi_{c}|},
\quad \text{and}\quad 
\tan\theta_{13}\approx \frac{\epsilon}{2\sqrt{2}}\frac{|1+a|}{\cos\frac{\phi_c}{2}}
$$
Both expressions, as expected, reduce to those in Eq.~(\ref{thetaatm}) for null phases. 
The small deviation of ATM mixing from $\pi/4$ suggests to  take $\phi_c\approx 0$, in which case one only has to replace $1\pm a$ by $|1\pm a|$ in our previous formulae.

Finally, Dirac CP phase can be directly calculated from $M_\nu$ by using the Jarlskog invariant~\cite{jarlskog}
$J =\sin2\theta_{12}\sin2\theta_{23}\sin\theta_{13}\cos\theta_{13}\sin\delta_{CP}/8$,
which can also be written as~\cite{branco}
\be
J = -
\frac{Im\left(H_{21}H_{32}H_{13}\right)}{\Delta m^2_{21}\Delta m^2_{31}\Delta m^2_{32}}~.
\ee
Hence, at order ${\cal{O}} (\epsilon^2)$, setting in the experimental values of neutrino oscillation parameters,  one gets
\be
\sin\delta_{CP}\approx -1.886~h_j~,
\ee
where, $h_j = 2\delta_b\epsilon|a|\sin(\phi_a +\phi_c) -(1+\delta_b\epsilon-2\delta_c\epsilon^2-|a|^2)\sin(2\phi_c) 
-|d|\epsilon^2 (\sin\phi_d+\sin(\phi_d+2\phi_c))$. 
This expression depends on the three free phases of $a$, $c$ and $d$. It vanishes when they are null, as it should be, and  it can easily accommodate  possible values for $\delta_{CP}$ within its expected experimental range. 

Numerically scanning the parameter space that is constrained by the condition to provide mass scales and mixings within their current 1$\sigma$ values, one gets the exact predictions for $\delta_{CP}$ depicted in figure 1, in terms of the main parameter associated to the breaking of $\mu-\tau$ symmetry, $a$. As it is notorious, a rather order one $a$ can actually account for a CP violating phase well within the present bounds at $3\sigma$ level. This indicates that a solution to neutrino oscillation parameters out of the proposed texture is possible without relaying on a strong breaking of the symmetry in the electron sector. All the points in the plot  correspond to relatively small values of $\epsilon$, within the range of $(0.2,0.35)$, and are consistent with order one $\delta_{b,c}$, within the range $(0.6,1)$. Also, they all were  found to have a value of $\phi_c$ very close to $0$, confirming the findings of our previous analytical analysis.

\begin{figure}[h]
\includegraphics[width=.5\textwidth]{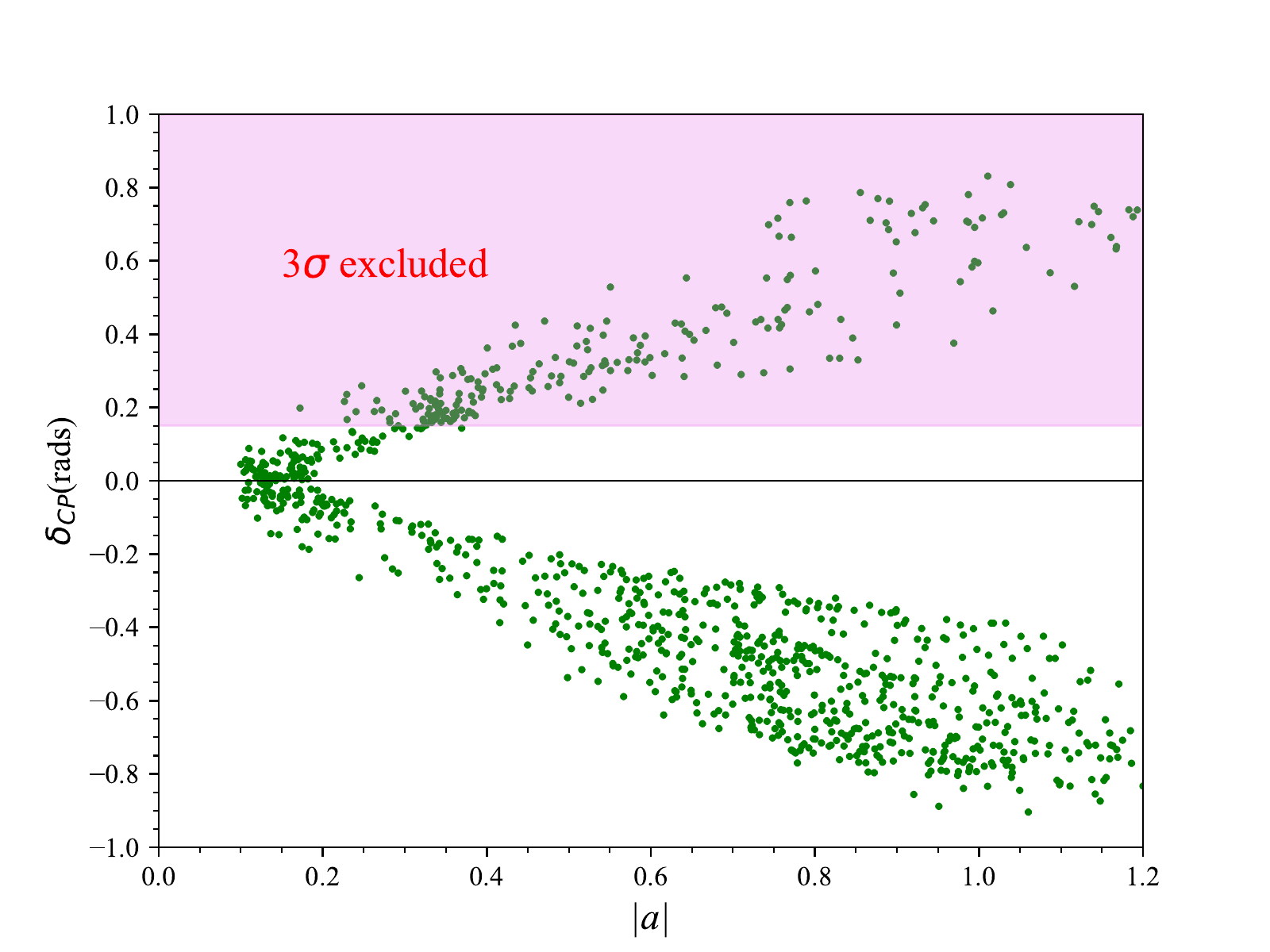}
\caption{CP violation phase predictions in terms of the $|a|$ parameter. The values currently excluded at $3\sigma$ level are within the shadowed region. }
 \label{delta3_CP}
\end{figure}

%%%%%%%%%%%%%%%%%%%%%%%%%%5
\section{Concluding remarks}
%%%%%%%%%%%%%%%%
In summary,  approximated $\mu-\tau$ symmetry in the neutrino sector appears as an interesting approach to the understanding of the way all neutrino oscillation parameters arise from neutrino mass matrix. As discussed above, in the exact symmetric limit, there is an appealing and general scenario where $\theta_{23}$ is maximal, and  $\theta_{13}$ is already non zero and small. Unlike many other approaches, in this scenario solar neutrino oscillation parameters emerge from the breaking of $\mu-\tau$ symmetry, which also contributes to fix other mixings to their observed values through out small corrections. The presented mass matrix realizes these predictions in the context of normal neutrino hierarchy using a single perturbation parameter, $\epsilon\sim\sqrt{\Delta m^2_{sol}./\Delta m^2_{ATM}}$ and may provide the right amount of CP violation.
One interesting implication of this is the fact that $|m_{ee}|\approx |d|\epsilon^2 m_0\sim 2|d|\times 10^{-3}~eV$, and thus, the expected amplitude for neutrinoless double beta decay is predicted to be small. A positive observation of such events in the ongoing experiments would rule out the proposed $M_\nu$. 
Furthermore, as the present analysis shows,  the proposed texture is stable under radiative corrections. Thus, its realization under a possible high energy flavor theory may also  deserve further attention.

Additionally, it is worth mentioning that the main observations regarding mixing outcomes from $\mu-\tau$ exchange symmetry made in here would also be reached for Dirac neutrinos, since in that case the calculation of the mixing matrix also uses an hermitian squared matrix, $H= M_DM_D^\dagger$, which has a similar form as the one given en Eq.~(\ref{H}) (see for instance Ref.~\cite{luna}).  In contrast,  these conclusions  do not hold for $\mu-\tau$ reflection symmetry~\cite{mtreflection,xing}
which adds CP conjugation to $\mu-\tau$ exchange, with Dirac neutrinos,  to predict non zero mixings without breaking the symmetry. A way to understand this is to notice that reflection symmetry do break the exchange symmetry through the phase difference among  $H_{e\mu}$ and $H_{e\tau} = H_{e\mu}^*$.
Of course, since switching off the phases trivially reduces one case into the other, the proposed connection of solar mixing to the origin of solar scale could still be realized in $\mu-\tau$ reflection symmetry models, interestingly, trough CP phases alone.

Finally, it is worth mentioning that there are also appropriate mass structures that work for inverted hierarchy, but they will be  presented in a forthcoming extended study. 

\vspace*{1ex}
%\begin{acknowledgments}
\section*{Acknowledgments}
Work partially supported by Conacyt, Mexico, under FORDECYT-PRONACES grant No. 490769.

%\end{acknowledgments}

\end{document}